\documentclass[twocolumn,showpacs,preprintnumbers,amsmath,amssymb]{revtex4}
\usepackage{graphicx}% Include figure files
\usepackage{dcolumn}% Align table columns on decimal point
\usepackage{bm}% bold math

\begin{document}

%\preprint{APS/123-QED}

\title{Coupling to a phononic mode in $\mathrm{Bi_{2-x}Pb_xSr_2CaCu_2O_{8+ \delta }}$: Angle-resolved photoemission}%}% Force line breaks with \\

\author{B. Ziegler, B. M\"uller, A. Krapf, H. Dwelk, C. Janowitz and R. Manzke}
 %\altaffiliation[Also at ]{Physics Department, XYZ University.}%Lines break automatically or can be forced with \\
%\author{Second Author}%
% \email{Second.Author@institution.edu}
\affiliation{
Humboldt-Universit\"at zu Berlin, Institut f\"ur Physik, Newtonstr. 15, 12489 Berlin
}

%\date{\today}% It is always \today, today,
             %  but any date may be explicitly specified

\begin{abstract}
The kink in the dispersion and the drop in the width observed by angle-resolved photoemission in the nodal direction of the Brillouin zone of $\mathrm{Bi_{2-x}Pb_xSr_2CaCu_2O_{8+ \delta  }}$ (abbreviated as (Pb)Bi2212) has attracted broad interest \cite{kink4,kink5,kink9,kink7, kord, kink10}. Surprisingly optimally lead-doped (Pb)Bi2212 with $\mathrm{T_C>89K}$ as well as the shadow band were not investigated so far, although the origin of the kink and the drop is still under strong debate. In this context a resonant magnetic-mode scenario and an electron-phonon coupling scenario are discussed controversially. Here we analyze the relevant differences between both scenarios and conclude that the kink and the drop are caused by a coupling of the electronic system to a phononic mode at least in the nodal direction. It is found that besides the dispersion and the drop in the width also the peak height as a new criterion can be used to define the energy scale of the interaction, giving a new means for a precise and consistent determination of the kink energy.
\end{abstract}

% PACS, the Physics and Astronomy
                             % Classification Scheme.
%\keywords{Suggested keywords}%Use showkeys class option if keyword
                              %display desired
\maketitle

\section{Introduction}
An important, but still open question is the coupling mechanism which causes high-temperature superconductivity.
To come closer to an answer concerning possible coupling scenarios it is important to find excitations which are due to an interaction of the electronic system to a collective mode and to search for its origin. If such a mode is strong enough and correlates with other superconducting properties then it could also be responsible for the pairing in high-temperature superconductivity. Therefore it is necessary to compare a large number of different high-temperature superconductors with the aim to find common properties in the electronic structure. Such a common property is the kink in the dispersion of the main CuO derived band in the nodal direction of the Brillouin zone. This kink is due to the coupling of the electrons to a collective mode and appears especially in Bi-cuprates with n=1, n=2, and n=3 $\mathrm{CuO_2}$ planes per unit cell. 
The bilayer components Bi2212 and (Pb)Bi2212 are the most frequently studied ones of the Bi-cuprates. The added lead suppresses the otherwise appearing $\mathrm{(\sim5x1)}$-superstructure, making the electronic structure of (Pb)Bi2212 more simple than that of Bi2212. Optimally lead-doped (Pb)Bi2212 with $\mathrm{T_c>89 K}$ has not been investigated so far concerning the kink in the nodal direction. 
Also the behavior of the shadow band regarding the kink was not considered up to now. Here we present a study of the kink in the dispersion and the drop in the width in nodal direction of optimally doped ($\mathrm{T_c=93 K}$) and slightly overdoped ($\mathrm{T_c= 83 K}$ and $\mathrm{T_c= 85 K}$) (Pb)Bi2212 samples by use of angle-resolved photoemission spectroscopy, where we not only consider the main, but also the shadow band. Additionally an optimally La-doped (Pb)Bi2201 sample ($\mathrm{T_c= 40 K}$) was used for reference.\\

\section{Experimental details}
The used high quality samples were grown out of a nonstoichiometric melt and were characterized by EDX, AC-susceptibilty, Laue diffraction, and LEED. The samples were investigated by use of angle-resolved photoemission spectroscopy at the Synchrotron Radiation Center in Madison/Wisconsin (USA). The measurements were done under UHV conditions with a Scienta SES2002 analyzer. Highest angle resolution of $\mathrm{0.1^\circ}$ and energy resolution of $\mathrm{\Delta E = 12.5 meV}$ has been achieved.
The polarisation of the radiation was parallel to the entrance slit of the detector. All measurements were done along the nodal direction, which means that the polarisation was parallel to $\mathrm{\Gamma X}$ (or $\mathrm{\Gamma Y}$).\

The high-temperature superconductivity of the Bi-cuprates is associated with the two-dimensional $\mathrm{CuO_2}$ planes, from where the signal near $\mathrm{E_F}$ originates. The photoemission intensity for a quasi-twodimensional system is given by
\begin{equation}
\label{I(kE)}
I(\vec{k},\omega)=I_0(\vec{k})f(\omega)A(\vec{k},\omega)
\end{equation}
Here $\vec{k}$ is the momentum parallel to the surface, $\omega$ is the energy of the initial state relative to the chemical potential, $f$ is the Fermi function, $I_{0}$ is proportional to the dipole matrix element $\mid M_{fi}\mid^2$ due to Fermis golden rule, and A is the one-particle spectral function. The experimental signal is a convolution of  $I(\vec{k},\omega)$ with the energy resolution and a sum over the momentum window plus an additive (extrinsic) background coming from secondary electrons.
The essential result in ARPES studies is the spectral function $A(\vec{k},\omega)$, which can be expressed by using the retarded Green's function $G(\vec{k},\omega)$
\begin{eqnarray}
\label{A(kE)}
A(\vec{k},\omega)&&=-\frac{1}{\pi}ImG(\vec{k},\omega)\nonumber\\
&&=\frac{1}{\pi}\frac{\mid\Sigma''(\vec{k},\omega)\mid}{[\omega-\epsilon_k-\Sigma'(\vec{k},\omega)]^2+[\Sigma''(\vec{k},\omega)]^2}
\end{eqnarray}
with the self-energy $\Sigma=\Sigma'+i\Sigma''$ and the bare dispersion $\epsilon_k$.
Furthermore, as can be seen by the generic expression for the spectral function $A(\vec{k},\omega)$ in Equation \ref{A(kE)} the peak position in an energy distribution curve (EDC) is determined by $\Sigma'(\vec{k},\omega)$ as well as $\Sigma''(\vec{k},\omega)$ because both terms are strongly energy dependent. On the other hand, if the self-energy $\Sigma$ is independent of k normal to the Fermi surface (and the matrix elements are a slowly varying function of k), then the corresponding momentum distribution curves (MDCs) are simple Lorentzians centered at \cite{Valla_Science99}
\begin{equation}
k = k_F + \frac{[\omega-\Sigma'(\omega)]}{\upsilon^0_F}.
\label{disp}
\end{equation}

This is obtained by approximating $\mathrm{\epsilon_k \simeq \upsilon^0_F (k-k_F) }$ in Equation \ref{A(kE)}. $\mathrm{\upsilon^0_F}$ represents the bare Fermi velocity normal to the Fermi surface.\cite{kink5}\

The full width at half maximum (FWHM) of the photoemission peak in a momentum distribution curve reflects the mean free path $\mathrm{l^\star = 1 / \Delta k}$ such that
\begin{equation}
\Delta k = \frac{\mid 2 \Sigma'' (\vec{k},\omega) \mid}{\upsilon^0_F}.
\label{width}
\end{equation} 
Kordyuk et al. \cite{kink7,kord} employ a self consistent determination of the real and the imaginary part of the spectral function with the bare Fermi velocity as input. But from Equations \ref{width}, \ref{disp}, and \ref{A(kE)} follows that it is necessary to know $\mathrm{\upsilon^0_F}$ or $\mathrm{\epsilon_k}$ ($\mathrm{\epsilon_k\simeq \upsilon^0_F(k-k_F)}$ for k near $\mathrm{k_F}$ \cite{kink5}) to get $\mathrm{\Sigma'}$ and/or $\mathrm{\Sigma''}$ from a fit of the photoemission spectrum. $\mathrm{\upsilon^0_F}$ or $\mathrm{\epsilon_k}$ can not be obtained directly from a photoemission spectrum. Nevertheless the general behavior of $\mathrm{\Sigma'}$ and $\mathrm{\Sigma''}$ is reflected in the dispersion and the FWHM curves, as Equations \ref{disp} and \ref{width} show.\

Generally the photoemission spectra, which are analyzed here, were fitted as follows. First the Fermi energy of a sample was determined by fitting a spectrum of a thin gold film evaporated on the sample. Then, for normalisation each spectrum was divided by such a spectrum of the sample covered with the gold film. The spectra were integrated over a window of 4meV at the energy axis and over a window of $\mathrm{0.28^\circ}$ at the angular axis.
From a complete spectra series the momentum distribution curves are derived as function of energy. These curves were fitted with a Lorentzian as indicated by Equation \ref{A(kE)}. From the Lorentzian the peak height at peak maximum, the k-position of the peak maximum and the full width at half maximum (FWHM) were taken for further consideration. 

\section{Results}

\begin{figure}%[ht]
	%\begin{center}
	\includegraphics{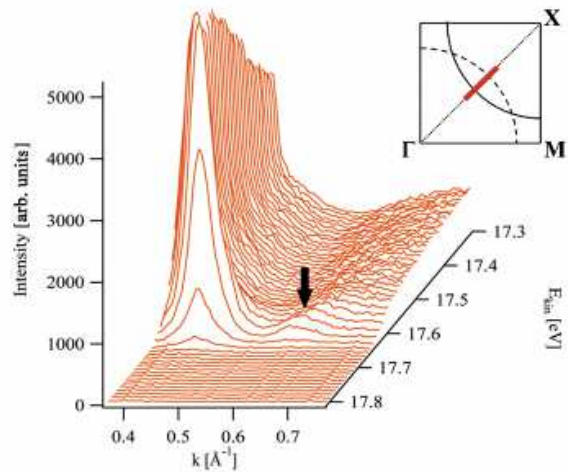} 
	%\end{center}
	\caption{Momentum distribution curves (MDCs) of optimally doped (Pb)Bi2212 (sample 1) taken at 20K with a photon energy of 22eV. The arrow indicates the shadow band. The red bar in the inset shows the measured direction of the Brillouin zone.}
	\label{sb}
\end{figure}

Fig. \ref{sb} shows typical momentum distribution curves, recorded in nodal direction at a temperature of 20K using a photon energy of 22eV. One can see the main band, which crosses the Fermi surface near $\mathrm{k=0.45\AA{}^{-1}}$, and the shadow band, which crosses the Fermi surface near $\mathrm{k=0.6\AA{}^{-1}}$ (see arrow).

Beside the kink in the dispersion of the main band also the behavior of the shadow band concerning the kink will be considered here. In a recent investigation of the dispersion of the main and the shadow band of optimally doped Bi2201, underdoped (Pb)Bi2212 and underdoped Bi2212 Koitzsch et al. \cite{Koitzsch_prb69} found that the dispersion of the main and the shadow band show a similar behavior. The same has been found for the FWHM regarding size and behavior. But this publication did not treat the kink explicitely.\

\begin{figure}
	\includegraphics[width=8cm]{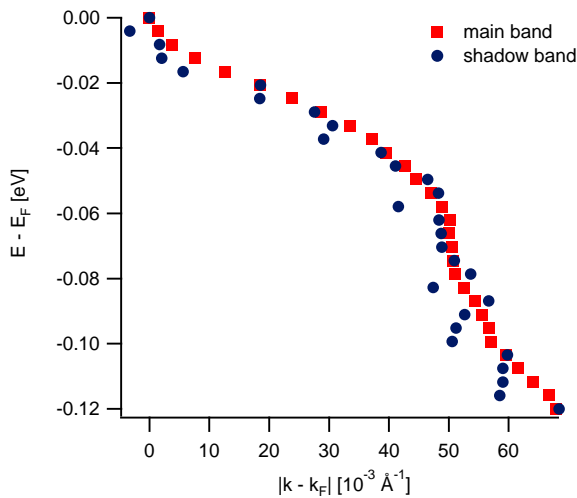} 
	\caption{Dispersion of the main (red squares) and the shadow band (blue dots) of optimally doped (Pb)Bi2212 (sample 1).}
	\label{HBSB}
\end{figure}

Fitting the spectra of optimally doped (Pb)Bi2212 as described in the previous section the following results were obtained.
Fig. \ref{HBSB} shows the dispersion of the main (red squares) and the shadow band (blue dots). A kink near $\mathrm{\hbar\omega \simeq 60meV}$ is clearly visible in the dispersion of both bands. In addition, both dispersions almost coincide which is to be expected if the shadow band is of structural origin as Mans et al. \cite{Mans} claim. Here we show that this is especially valid for the kink in the dispersion in the nodal direction.  \

\begin{figure}
	\includegraphics[width=7cm]{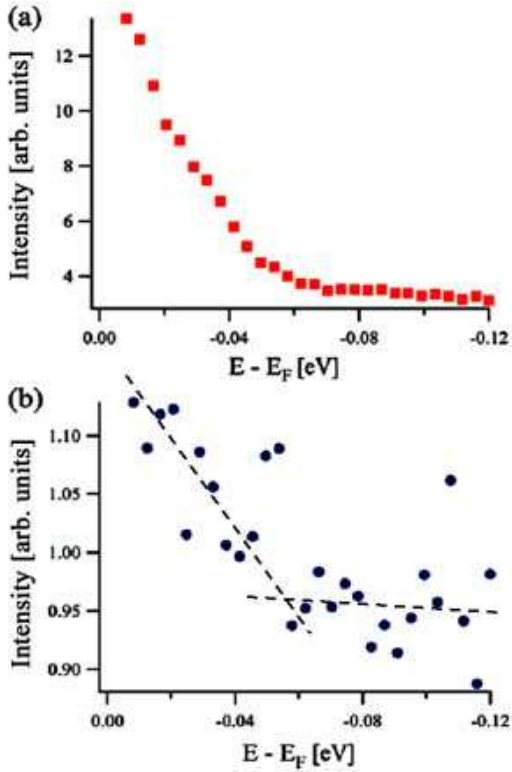} 
	\caption{Intensity at peak maximum of the main band (a) and of the shadow band (b) of optimally doped (Pb)Bi2212 (sample 1). The black dashed lines are guides to the eye.}
	\label{HBSBImax}
\end{figure}

Moreover it can be observed that besides the kink in dispersion there is also a "kink in intensity" or more precisely a "kink in peak height". Additional to the commonly used criteria dispersion and width also the peak height shall be discussed as an additional and alternative means to define the energy position of the kink. Fig. \ref{HBSBImax}(a) shows a plot of the MDC peak height of the main band at the maximum of the peak as a function of $\mathrm{\hbar\omega = E - E_F}$. One can clearly see a sharp kink near 60meV, which corresponds to the position of the kink in dispersion (see Fig. \ref{HBSB}). Due to the restricted statistics a less sharp "kink in peak height" is also visible in Fig. \ref{HBSBImax}(b) for the shadow band.

\begin{figure}[h]
	\includegraphics[width=8cm]{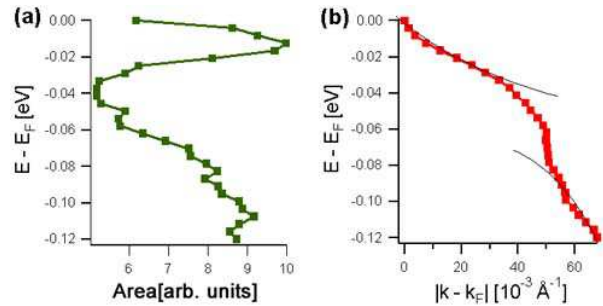} 
	\caption{(a) Area under the momentum distribution curves of optimally doped (Pb)Bi2212 (sample 1) showing a minimum between $\mathrm{\hbar\omega \simeq 30meV}$ and $\mathrm{\hbar\omega \simeq 60meV}$. (b) Dispersion of the main band of optimally doped (Pb)Bi2212 (sample 1). The black lines are guides to the eye.}
	\label{Fl}
\end{figure}

The kink in peak height, which was analyzed here for the first time, 
improves the ability to determine the precise position of the kink. A knowledge of the precise position of the kink is important to distinguish between the different possible coupling modes, which are located at very similar energies (for details see \ref{disc}). 
The kink in the peak height can qualitatively be understood in connection with a shift of spectral weight. Fig. \ref{Fl}(a) shows the area under the momentum distribution curves as a function of $\mathrm{\hbar\omega = E - E_F}$. For comparison the corresponding dispersion is plotted in Fig. \ref{Fl}(b). In agreement with theoretical calculations, which predict a splitting of the band into two branches at the energy of the coup-ling mode \cite{Norman}, one can interpret the minimum between $\mathrm{\hbar\omega \simeq 30 meV}$ and $\mathrm{\hbar\omega \simeq 60 meV}$ in Fig. \ref{Fl}(a) as the region of low spectral intensity between the two split bands. This minimum and the kink in the peak height indicate that spectral weight is shifted to lower binding energies and confirms the predicted two-branch behavior.
\

\begin{figure}[h]
	\includegraphics[width=7cm]{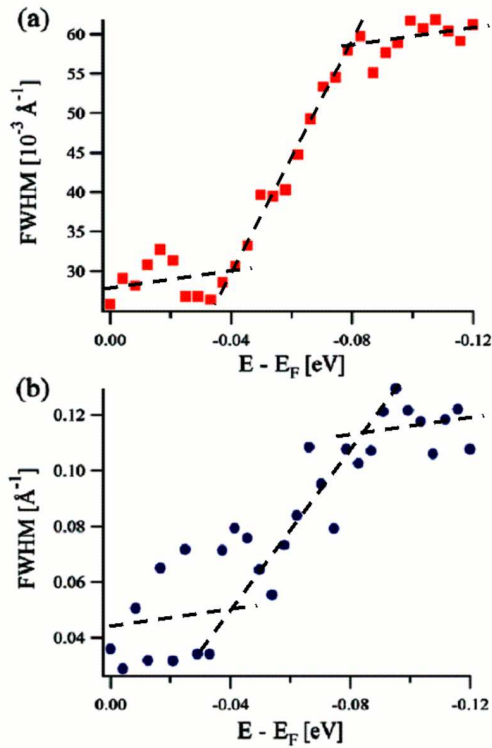} 
	\caption{FWHM of optimally doped (Pb)Bi2212 (sample 1) of the main band (a) and of the shadow band (b). The black dashed lines are guides to the eye.}
	\label{HBSBB}
\end{figure}

The interpretation of the FWHM curves, which are plotted in Fig. \ref{HBSBB}(a) and Fig. \ref{HBSBB}(b), is less straight forward. Around 60meV, where the kink in the dispersion and the kink in peak height are situated, a drop occures. The general line shape of the FWHM curve is consistent with published results of other (Pb)Bi2212 and Bi2212 samples (see for example \cite{kink4,kink5,kink9,kink7,kink10}). But there is no \linebreak general agreement about the determination of the position of the so called "scattering rate kink" in the literature. Kordyuk et al. \cite{kink7} usually take the onset of the drop as kink position, but then the scattering rate kink in \cite{kink7} is located around $\mathrm{\hbar\omega \simeq 100meV}$ and not at the same position as the kink in the dispersion. Koitzsch et al. \cite{kink10} did not discuss the position of the "scattering rate kink". For nearly optimally doped (Pb)Bi2212 ($\mathrm{T_c = 89K}$) they found a kink in dispersion around $\mathrm{\hbar\omega \simeq 60meV}$ and a drop in the scattering rate with an onset near $\mathrm{\hbar\omega \simeq 80meV}$ and an offset near $\mathrm{\hbar\omega \simeq 40meV}$. Optimally doped (Pb)Bi2212 shows the same behavior (see Fig. \ref{HBSB} and Fig. \ref{HBSBB}(a)). This result implies, that the "scattering rate kink" is located at half height between onset and offset of the drop. The drop can be explained as an abrupt change of the mean free path $\mathrm{l^{\star}\sim 1/ \Sigma''}$ (see Equation \ref{width}) at the energy of the kink in the dispersion and the kink in the peak height.

From a comparison of the FWHM curves in Fig. \ref{HBSBB}(a) and Fig. \ref{HBSBB}(b) with the corresponding dispersion and peak height curves (see Fig. \ref{HBSB} and Fig. \ref{HBSBImax}), it can be concluded that the "scattering rate kink" is located at half height between onset and offset of the drop. So "drop" instead of "kink" characterizes more precisely the here observed phenomenon.\

Also concerning the general line shape of the FWHM curves of the main and the shadow band behave similarly although a kink is hardly visible in Fig. \ref{HBSBB}(b) due to the restricted statistics. Albeit small deviations the FWHM near $\mathrm{E_F}$ is the same in both bands like also found in \cite{Koitzsch_prb69, Mans}, but increases for higher binding energies. \

\begin{figure}[h]
	\includegraphics[width=8cm]{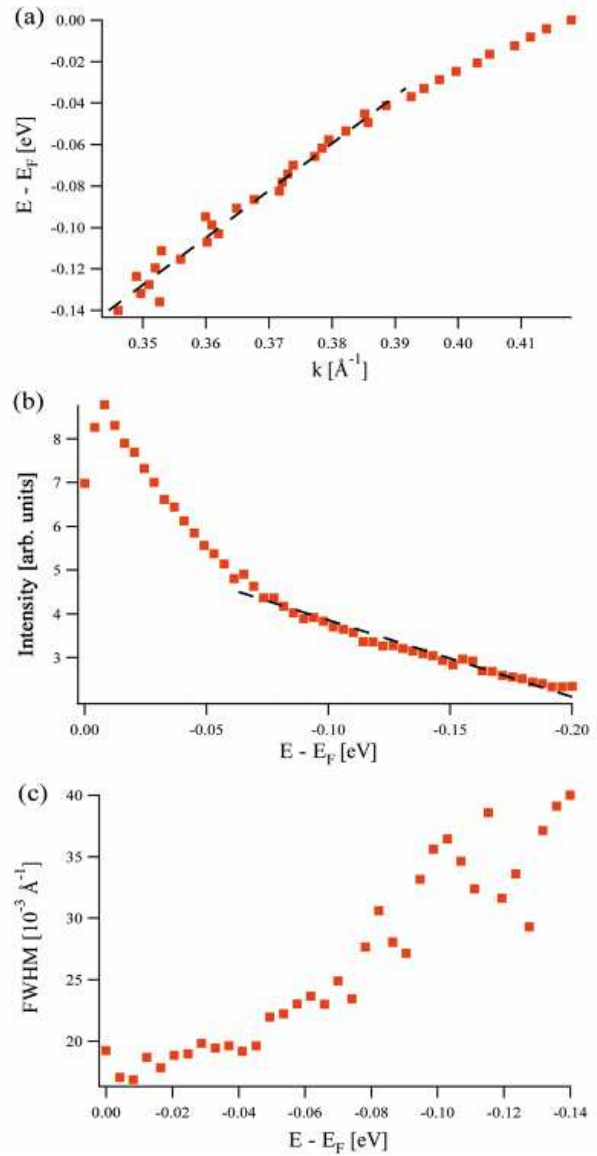} 
	\caption{(a) dispersion, (b) peak height, and (c) FWHM of the main band of optimally doped (Pb)Bi2201 (sample 4); red squares: data, black dashed line: guide to the eye.\label{n1}}
	
\end{figure}

Analogous results were obtained for the (Pb)Bi2201 sample although the features appear less pronounced. This is shown in Fig. \ref{n1} where the kink in the dispersion, the kink in the peak height, and the drop in the FWHM of the main band appear near $\mathrm{\hbar\omega \simeq 60meV}$.\

Table \ref{kink} shows the different samples studied, their transition temperature and Pb content, the number of analyzed spectra series and the obtained kink energy. No general trend for a scaling of the kink energy with $\mathrm{T_c}$ is observable. The averaged position of the kink, taking into account all analyzed spectra of (Pb)Bi2212 samples, is at $\mathrm{\hbar\omega_{kink} = (64 \pm 6) meV}$.\ 

\begin{table}[h]
\begin{center}
\begin{tabular}{|c|c|c|c|c|c|}
\hline
sample & $\mathrm{T_c [K]}$ &  doping & x & number of &  $\mathrm{\hbar\omega_{kink} [meV]}$\\
&&&& spectra series &\\
\hline
1 & 93 & opt. ($\mathrm{O_2}$) & 0.28 & 3 & $\mathrm{58 \pm 6}$\\
2 & 85 & overd. ($\mathrm{O_2}$) & 0.27 & 1 & $\mathrm{69 \pm 5}$\\
3 & 83 & overd. ($\mathrm{O_2}$) & 0.27 & 6 & $\mathrm{65 \pm 5}$\\
4 & 40 & opt. (La)& 0.43 & 1 & $\mathrm{60 \pm 8}$\\
\hline
\end{tabular}
\end{center}
\caption{The measured three different (Pb)Bi2212 samples 1-3 and the (Pb)Bi2201 sample 4, their $\mathrm{T_c}$, doping level, Pb content x, the number of the analyzed spectra series, and the kink energy $\mathrm{\hbar\omega_{kink}}$.}
\label{kink}
\end{table}

\begin{figure}
	\includegraphics[width=8cm]{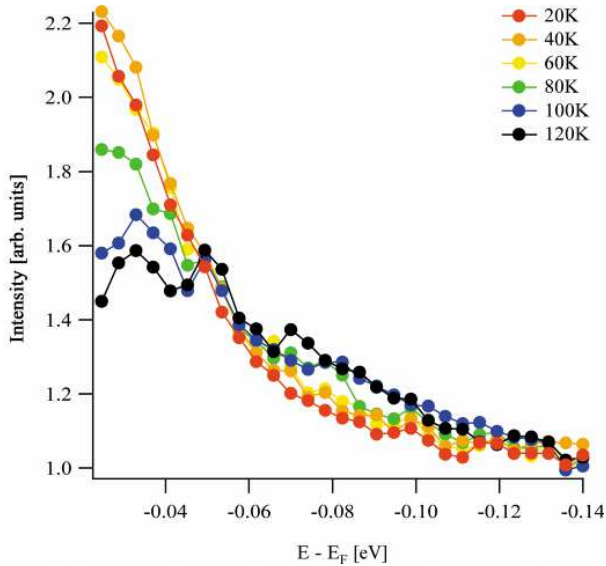} 
	\caption{Intensity at peak maximum for different temperatures of slightly overdoped (Pb)Bi2212 (sample 3).}
	\label{TImax}
\end{figure}

Measurements at different temperatures between 20K and 120K were performed at an slightly overdoped (Pb)Bi2212 sample ($\mathrm{T_c = 83 K}$, sample 3) using a photon energy of 22eV. 
In agreement with former published literature the kink in the dispersion persists at least up to $\mathrm{T=120K}$ and sharpens with decreasing temperature \cite{kink4,kink5}. The drop in the width vanishes between 100K and 120K and is presumably related to the pseudogap temperature $\mathrm{T^\star}$ and not $\mathrm{T_c}$. In Fig. \ref{TImax} the peak height at peak maximum is plotted for different temperatures. A closer look shows that the kink in the peak height around 65meV gradually vanishes with increasing temperature and also disappears almost at 120K. This indicates that the appearence of kink is not connected with $\mathrm{T_c}$ but presumably with $\mathrm{T^\star}$.\

\section{Discussion}\label{disc}
A coupling of the electronic system to a collective mode generally appears in photoemission data as a kink in the dispersion, a kink in the peak height and a drop in the width. If one compares these three features as criteria for a determination of the mode energy then one has to realize that the kink in the peak height allows a very precise determination while the drop in the width is an unsure indicator (see especially Fig. \ref{n1}). Together with the kink in the dispersion the kink in the peak height is a strong criterion for the determination of the energy where an interaction with a collective mode takes place. Since the angular scans only cover a very limited angular range we expect the matrix element variations to be of minor importance for the peak height criterion.\ 

Provided that the kink energy is a manifestation of the most prominent interaction of the electronic degrees of freedom with a boson we concentrate on its discussion. At present time two different coupling mechanisms are controversially discussed in the literature. The first one is based on the resonant magnetic-mode scenario and the second one on the electron-phonon coupling scenario. Two reasons complicate a decision between these two. At first, the general manifestation of a coupling mode in ARPES spectra without further information does not allow one to decide between a phononic or a magnetic mode. In both cases the spectra would show a kink in the dispersion. At second, for Bi-cuprates the magnetic and phononic modes are incompletely investigated.\

Magnetic resonance modes can generally be observed in inelastic neutron scattering (INS). But there are no published results of measurements on (Pb)Bi2212 and only a few on Bi2212 because the crystals typically grow as thin plates with volumina much too small for INS. The magnetic excitations in Bi2212 \cite{Fong99,He} were observed at energies around $\mathrm{E_{res} = 40 meV}$ scaling with $\mathrm{T_c}$. A remnant of these modes persisted far into the pseudogap state. Further, a comparison of the results obtained from Bi2212 with those from $\mathrm{YBa_2Cu_3O_{6+x}}$ shows that the energy of the magnetic resonance mode scales with $\mathrm{T_c}$ in both the underdoped and the overdoped regimes \cite{He,Dai}. Recently an additional magnetic mode was observed in $\mathrm{YBa_2Cu_3O_{6+x}}$ at $\mathrm{E_{res} = 57meV}$ \cite{Eremin}.\

Also phononic modes can be measured by inelastic neutron scattering (INS). For the above mentioned reasons there are no publications concerning (Pb)Bi2212 or Bi2212. But in $\mathrm{YBa_2Cu_3O_{6+x}}$ phononic modes were observed at $\mathrm{E_{res} = 55meV}$ and $\mathrm{E_{res} = 75meV}$ independent of $\mathrm{T_c}$ \cite{Petrov}.\ 

There are a lot of arguments in favour of both scenarios. So it is necessary to search for the differences between the two. First, it has to be realized that for both the coupling and their manifestation as kinks in the photoemission spectra persist presumably up to $\mathrm{T^\star}$. Also the energies where coupling modes or a kink in the photoemission spectra can be expected are similar, between 40meV and 75meV. The main difference is the observation that the energy of the magnetic resonance mode scales with $\mathrm{T_c}$ in contrast to the energy of the phononic modes.
This means for ARPES measurements that the difference between magnetic and phononic coupling is the scaling of the kink energy with $\mathrm{T_c}$ in the magnetic case in contradiction to phononic coupling. The kink energies of the different samples are listed in Table \ref{kink}. For the three measured (Pb)Bi2212 samples no scaling of the kink energy with $\mathrm{T_c}$ is observed. Also former published measurements (see for example \cite{kink4,kink7, kink10, Sato}) covering a wide doping range did not show a dependence of the kink energy on $\mathrm{T_c}$. This result argues against a coupling to the magnetic resonance mode. Furthermore it is consistent with results of photoemission measurements on $\mathrm{La_{2-x}Sr_xCuO_4}$ (LSCO) in nodal direction, where also no dependence of the kink energy on doping was found for a wide doping range \cite{Zhou}. 
Additionally the nodal kink in the dispersion is independent of the number of $\mathrm{CuO_2}$ layers in the unit cell \cite{Sato}. This means that the energy of the kink is independent of $\mathrm{T_c}$, what is also an argument in favour of a phononic coupling as well as the fact that the kink in dispersion in the nodal direction persists even above $\mathrm{T^\star}$.  \

\section{Summary}
Along the nodal direction all samples, especially the optimally doped samples, showed a kink in the dispersion of the main band.
Around the position of the kink in the dispersion curve at $\mathrm{\hbar\omega_{kink} = (64 \pm 6) meV}$ appeared a drop in the full width at half maximum (FWHM) curve. This means that above $\mathrm{\hbar\omega_{kink}}$ the electronic system couples very effectively to a collective mode and is therefore strongly damped. Additionally, we found a kink in the peak height, which was not analyzed up to now. This kink in the peak height appeared at the same position as the kink in the dispersion and the drop in the width and allows a more precise determination of the kink position. It can be explained as an indicator for a shift of spectral weight to lower binding energies.\

The shadow band behaved analogously concerning the kink in the dispersion curve and the drop in the FHWM curve. Because the intensity of the shadow band is much lower than the intensity of the main band it was not possible to determine a kink in the peak height also for the shadow band.\

In agreement with former published literature the kink in dispersion appears independently of temperature and only sharpens at lower temperature. The drop in the FWHM curves persists at temperatures higher than $\mathrm{T_c}$, but disappears near $\mathrm{T^\star}$. The latter is also valid for the kink in peak height, what shows that the observed kink in the peak height is closely connected with the kink in the dispersion and the drop in the width.\

Possible reasons for the kink in the peak height, the kink in the dispersion, and the drop in the width are coupling to a magnetic or to a phononic mode. The main difference between these two scenarions is the scaling of the magnetic mode energy with $\mathrm{T_c}$. But such a scaling of the energy of the kink could not be observed here and also not in the published literature. Therefore it is very likely that phononic coupling is responsible at least for the kink in the nodal direction.

\section{Acknowledgement}
We gratefully thank the staff of the Synchrotron Radiation Center in Madison/Wisconsin (USA) for the excellent support during the measurements and the DFG for financial support. This work is based upon research conducted at the Synchrotron Radiation Center, University of Wisconsin-Madison, which is supported by NSF under Award No. DMR-0084402.

%\bibliography{paper}% Produces the bibliography via BibTeX.

\end{document}